\documentclass{vgtc}                          % final (conference style)
\graphicspath{{figures/}{pictures/}{images/}{./}} % where to search for the images

\usepackage{times}                     % we use Times as the main font
\usepackage{tabu}                      % only used for the table example
\usepackage{booktabs}                  % only used for the table example
\usepackage{lipsum}                    % used to generate placeholder text
\usepackage{mwe}                       % used to generate placeholder figures

\usepackage{mathptmx}                  % use matching math font

\onlineid{0}

\vgtccategory{Research}

\vgtcinsertpkg

\title{The Hall of Singularity: VR Experience of Prophecy by AI}

\author{Jisu Kim\thanks{e-mail: fh17jsk@gmail.com }\\ %
        \scriptsize Department of Art \& Technology, Sogang University, Seoul, South Korea %
\and Kirak Kim\thanks{e-mail: kirak@sogang.ac.kr}\\ %
     \scriptsize Department of Art \& Technology, Sogang University, Seoul, South Korea %
     }

\abstract{
    ‘The Hall of Singularity’ is an immersive art that creates personalized experiences of receiving prophecies from an AI deity through an integration of Artificial Intelligence (AI) and Virtual Reality (VR). As a metaphor for the mythologizing of AI in our society, ‘The Hall of Singularity’ offers an immersive quasi-religious experience where individuals can encounter an AI that has the power to make prophecies. This journey enables users to experience and imagine a world with an omnipotent AI deity. % filler text. Replace with your abstract.
} % end of abstract


\begin{document}

%% The ``\maketitle'' command must be the first command after the
%% ``\begin{document}'' command. It prepares and prints the title block.

%% the only exception to this rule is the \firstsection command
\firstsection{Introduction}

\maketitle

%% \section{Introduction} %for journal use above \firstsection{..} instead
‘Technological Singularity’ is a quasi-religious hypothesis about a certain future point in time when artificial intelligence (AI) surpasses human intelligence and capabilities\cite{shanahan2015technological}. Mythologizing AI prevails with various media platforms especially in the form of comparing future abilities of AI to those of omnipotent and omniscient deities, or monotheistic god.

Untransparent characteristics of Large AI models also plays a significant role in reinforcing this myth. From the Delphic Oracle to Catholicism, and Korea's Mudang (shamanistic medium), we individuals have always relied on so called “Mediums(an individual held to be a channel of communication between the earthly world and a world of spirits)\cite{medium}’ to connect with deities or divine beings, receiving prophecies and messages on their behalf\cite{yang1988korean}. Similarly, the limited accessibility of large AI models, which necessitates reliance on intermediary programs for non-AI professionals, contributes to the perpetuation of mythical misconceptions surrounding AI. Non-Ai professionals rely on medium programs like OpenAI playground that connect individuals with the data they want and deliver the results\cite{openai}. This limited accessibility of large AI models enhances the mythological misconceptions surrounding AI since it is  only reachable and understood by AI professionals and medium programs.

In this work, we attempt to portray the prevailing state of AI mythologization in our society through this metaphor:  ‘The medium of artificial intelligence that possesses omnipotent power like deity’. We compared utilization of large AI models via intermediary programs with the act of entrusting and relying on mediums that connect with deity. This metaphor effectively captures the essence of the situation where individuals depend on intermediary programs to gain access to the large AI models without clear understanding.

Moreover, we used Virtual Reality as a medium for accessing large artificial intelligence models. Recent large artificial intelligence models like Stable Diffusion and GPT have gained significant attention as a tool to improve user engagement\cite{cao2023comprehensive}. However, the majority of users still utilize large artificial intelligence models within the framework of traditional computer user interfaces (ex: websites). We aimed to provide a more immersive experience for users by using VR to mediate Artificial Intelligence. Designing VR levels more appropriately, we managed the drawbacks of using artificial intelligence model such as latency in processing generation in AI.

\section{User Walkthrough}
\begin{figure}[ht!]
    \centering
    \includegraphics[width=0.4\textwidth]{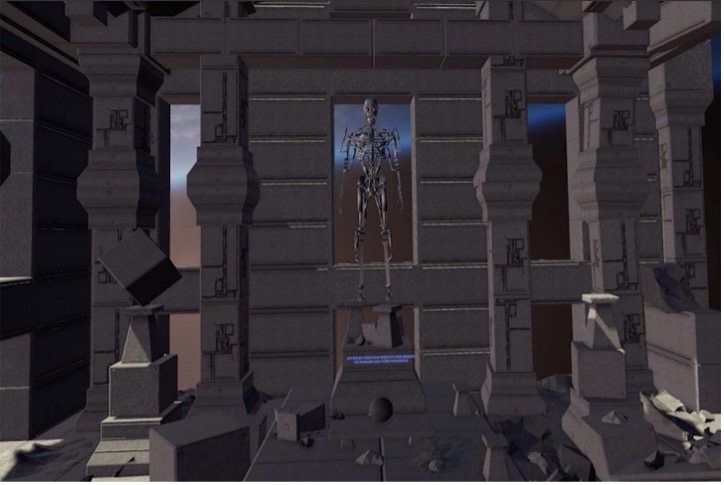}
    \caption{‘The Medium of AI deity’ in the virtual temple. User poses questions about the future in front of the medium.}
    \label{fig:figure1}
\end{figure}
1. User enters the Hall of Singularity in Virtual Reality through HMD headset. The temple, where they can encounter the "Medium of AI deity," engulfs users with its reminiscence of celestial space.

2. User can spacewalk in a cosmic temple via swimming motion by using hand controllers. This motion synchronizes physical movement with virtual movement and helps user to navigate inside the temple.

3. Once user encounter the ‘Medium of the AI deity’, they are guided to ask one question to the Medium using their own voices, regardless of the length of their inquiries.

4. Before the prophecy delivered, visitors can explore the temple's ceiling and floor, where they can wander and observe prophecy videos from previous users.

5. When the prophecy is ready, user is teleported to the space where they can watch a 30-second video showcasing their prophecy. 

\section{AI – VR Integration}
\begin{figure}[ht!]
    \centering
    \includegraphics[width=0.4\textwidth]{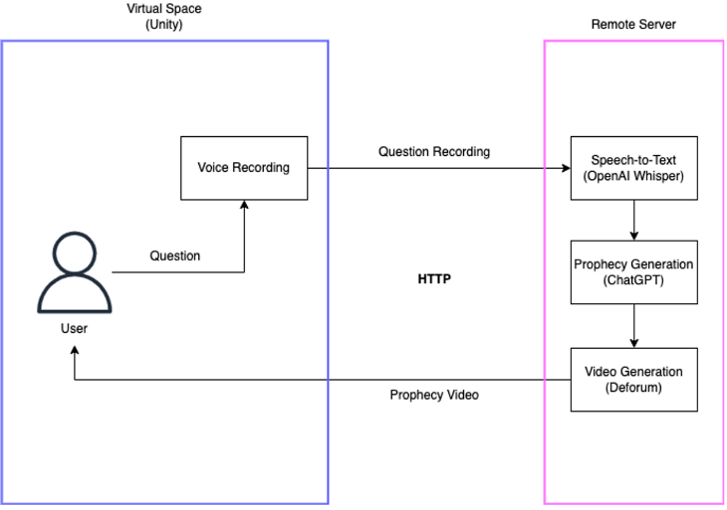}
    \caption{‘The exchange of information between the virtual space and the remote server. Within the virtual space, user can record their voices, while on the remote server, Python code is executed to run the artificial intelligence models. The connection between the two is established through HTTP.}
    \label{fig:figure2}
\end{figure}
To deliver mysterious and personalized experience to user, various Large artificial intelligence models were used throughout the entire process of the work. We used Whisper model from Open AI to enable user to speak directly to the ‘Medium of deity’, and ChatGPT to generate personalized prophecy that responds user’s questions. And we used Deforum model to deliver the prophecy into video format to infuse more ambiguity and allow various interpretations from the user.
\subsection{Real-time Voice Recognition and Translation through Whisper Model}
In the first stage, we used Whisper for speech-to-text and its real-time translation function to convert the voice data from user into English at the same time since the AI models such as ChatGPT and Deforum provide better quality of output in English. Due to the use of Whisper's translation capabilities, users can ask questions in various languages, including Korean, as demonstrated in the demo video. The voice data spoken by user is converted into English text and the result is sent to the input prompt of ChatGPT.
\subsection{Prophecy Generation with ChatGPT}
In the second stage, we used ChatGPT model to generate a prophecy that responds to user’s question. We gave predefined few shot examples to ChatGPT to generate a prompt for Deforum model that helps it to make consistent style of video. After Whisper model finishes the translation, the result is sent and added to the input of the predefined prompt in ChatGPT. Then, the ChatGPT model generates the text version of prophecy based on the input text. 
\subsection{Video Prophecy Generation with Deforum}
We decided to present the prophecy into Video format to allow diverse interpretation in the prophecy. We transformed prophecy text from ChatGPT into Video through Deforum model that enables ‘Text to Video” generation. It takes 2 minutes to generate 30 seconds of video.
\subsection{Integrating AI and VR through HTTP Communication}
The connection between large artificial intelligence models and the virtual space is established through HTTP communication. Artificial intelligence models are accessed through APIs on a remote server, and the virtual space communicates with the remote server through HTTP requests, exchanging information. In this system, User’s voice data is sent from virtual space to remote server, and generated prophecy data is sent from remote server to virtual space.
\section{Designing Virtual Reality}
\begin{figure}
    \centering
    \includegraphics[width=0.4\textwidth]{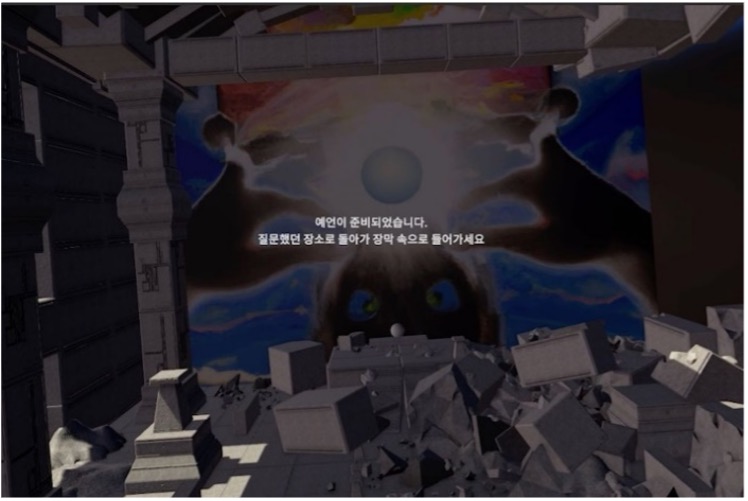}
    \caption{‘The veil conceals the medium during the generation of prophecy. The illustration on the veil portrays the depiction of the deity interpreted by the large AI models (GPT and DALL-E2).}
    \label{fig:figure3}
\end{figure}
The background of ‘The Hall of Singularity’ is a cosmic temple. Spacewalk is implemented via swimming motion using hand controllers. User can freely navigate through the space by holding the controllers and doing swimming strokes. We chose swimming as the method of locomotion in order to enhance presence. Swimming as spacewalk is well-suited for the cosmic background.  Also, user’s movements in the virtual space are synchronized with the real-world movements while swimming. Synchronizing user's real-world movements with their movements in the virtual space enhances presence for user\cite{lee2017study}. The 3D modeling objects in the temple are designed to have larger scale than objects in our daily life.  It makes user feel overwhelmed and gives a sense of awe. Since swimming is used for locomotion, user is free to move both horizontally and vertically as well to touch the ceiling.

However, when user pose questions, they are guided to ask one question from the floor. This guidance makes user to look up ‘The Medium of the AI deity’ while they ask for prophecy. To enhance presence in Virtual Reality, we enabled interaction through voice by using Whisper. This function allows user to ask questions without the need for a separate virtual keyboard. 

The prophecy generation begins after user asked their question. The Medium of the deity is concealed behind the veil. It takes approximately 2 minutes for the AI models to generate prophecy video. During this generation time, user is allowed to explore and watch various prophecy videos from previous user. The videos are placed all over the temple from the ceiling to the bottom. These videos are strategically placed in various locations. With their swimming motion, this design provides a multidirectional viewing experience.

Once the generation is complete, the color of the veil changes. As user approaches the veil, they are transported to a separate enclosed space where they can view the prophecy.

\begin{figure}
    \centering
    \includegraphics[width=0.4\textwidth]{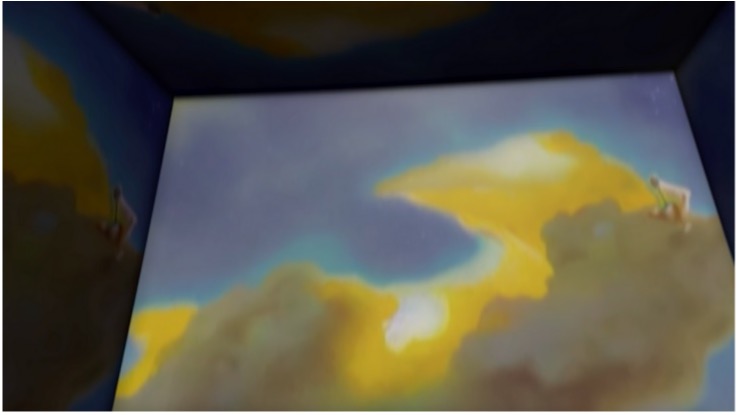}
    \caption{‘User watches the prophecy within an enclosed space. The prophecy is displayed on all the walls, with post-processing effects like motion blur applied.}
    \label{fig:figure4}
\end{figure}

The prophecy videos are projected onto all surfaces of the square space. Post-processing effects like motion blur infuse a unique visual atmosphere and creates a sense of detachment from previous experiences. Once the prophecy video ends,  user is transported back to the starting point where they can start another journey to ask additional questions. This effect allows a user to continue their exploration and ask further inquiries within the virtual space.
Here, variation in background music and ambience are also implemented to make the transition between each stage more noticeable. 

\section{Conclusion}
"The Hall of Singularity" invites user to explore the realm of contemporary artificial intelligence technology, offering a quasi-religious experience as metaphor of mythologization around AI. In this immersive art, user can experience a virtual world where artificial intelligence possesses omnipotent power as it is portrayed through mass media.

Generating prophecy, user’s voice data converts into English text with Whisper model’s real-time voice recognition and translation. Then, ChatGPT generates a prophecy based on user queries, with the input text incorporating the translated voice data and predefined prompts. This mechanism  ensures producing coherent results  regardless of the content of input(question) from user. Lastly,  Deforum model transforms the prophecy text into a video format. This abstract form of prophecy allows user for richer and diverse interpretations.

"The Hall of Singularity" is an experimental attempt that use VR as a new medium utilizing large AI models. Designing VR levels for AI models addressed drawbacks of large AI models, such as latency of generation time. Incorporating large AI models in a virtual space offered distinctive advantages of infusing dynamics into VR environment. This exploration paves the way for a more interactive and engaging user experience at the intersection of VR and AI. 
\acknowledgments{The authors would like to thank Prof. Dasaem Jeong at Sogang University, for instructing AI \& Creativity course (ANT6040, Fall 2022) where the subject idea of this artwork were begun. We would also like to thank to Heeyeon Yeon, a master’s student at department of artificial intelligence of Sogang University, for providing technical assistance.}

\bibliographystyle{abbrv-doi-hyperref}

\bibliography{template}
\end{document}